\documentclass[pss,fleqn]{w-art}
\usepackage{times}
\usepackage{w-thm}
\theoremstyle{plain}

\theoremstyle{definition}

\usepackage[]{graphicx}
\chardef\bslash=`\\ % p. 424, TeXbook

\begin{document}
\renewcommand{\copyrightyear}{2006}
\DOIsuffix{theDOIsuffix}
\Volume{XX}
\Issue{1}
\Month{01}
\Year{2006}
%%    First and last pagenumber of the article. If the option
%%    'autolastpage' is set (default) the second argument may be left empty.
\pagespan{1}{}
\Receiveddate{10 January 2006}
\Reviseddate{}
\Accepteddate{}
\Dateposted{}

\keywords{Ising model, eight-vertex model, non-universal criticality,  bicritical points.}
\subjclass[pacs]{05.50.+q, 05.70.Jk, 75.10.Hk, 75.40.Cx, 68.35.Rh}

%% \pretitle{Editor's Choice}

\title[Weak universal critical behaviour]{Weak universal critical behaviour
of the mixed spin-(1/2, $S$) \\ Ising model on the union jack (centered square) lattice: \\
integer versus half-odd-integer spin-$S$ case}

\author{Jozef Stre\v{c}ka\footnote{E-mail: {\sf jozkos@pobox.sk}, Phone: +421\,55\,6222121\,no.231,
           Fax: +421\,55\,6222124}, Lucia \v{C}anov\'a and J\'an Dely} 
\address[]{Department of Theoretical Physics and Astrophysics, 
Faculty of Science, \\ P. J. \v{S}af\'arik University,  
Park Angelinum 9, 040 01 Ko\v{s}ice, Slovak Republic}

%%    \dedicatory{This is a dedicatory.}

\begin{abstract}

The mixed spin-(1/2, $S$) Ising model on the union jack (centered square) lattice is investigated 
by establishing the mapping relationship with its corresponding eight-vertex model. An interplay 
between the nearest-neighbour interaction, the competing next-nearest-neighbour interaction 
and the single-ion anisotropy gives rise to a rather complex critical behaviour displayed in 
the reentrant phase transitions, the weak universal critical behaviour, as well as, a presence 
of first- and second-order phase transitions. The most interesting finding to emerge from the 
present study relates to a variation of the weak-universal critical exponents along the line 
of bicritical points, which is being twice as large for the mixed spin-(1/2, $S$) systems with 
the integer spin-$S$ atoms as for the ones with the half-odd-integer spin-$S$ atoms.

\end{abstract}

\maketitle                  

\section{Introduction}

One of the most fascinating exactly solvable lattice-statistical models exhibiting cooperative spontaneous order and extraordinary rich critical behaviour is being the \textit{eight-vertex model}. Over the last three decades this model turned out to have various applications in seemingly diverse research areas, actually, its numerous special cases proved their usefulness as the ice models, the ferroelectric KDP and F models, 
the free-fermion models, the dimer models, or the familiar planar Ising models originally introduced for insulating magnetic materials \cite{Bax71}. It is worthwhile to mention, moreover, that Baxter's exact solution of the symmetric (zero-field) eight-vertex model \cite{Bax82} has led to a remarkable breakthrough 
in the phase transition theory, since it provided a convincing evidence for a continuous change of critical exponents evidently contradicting the universality hypothesis \cite{Gri70}. Exact results reported 
on the continuously varying critical exponents consecutively inspired Suzuki in order to propose the weak universality hypothesis \cite{Suz74}, which allows changes of the critical exponents that do not violate 
the concept based on scaling laws \cite{Wid65}. 

\begin{figure}[htb]
\vspace{0.0cm}
\begin{center}
\includegraphics[width=5cm]{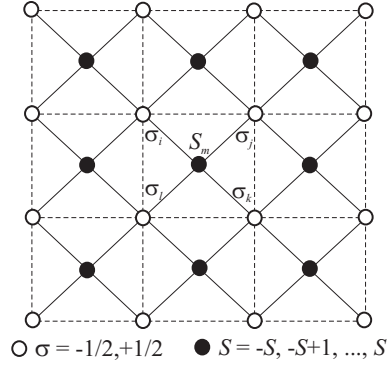}
\end{center}
\vspace{-0.2cm}
\caption{Schematic representation of the mixed spin-1/2 and spin-$S$ Ising model on the union jack (centered square) lattice. Solid (broken) lines depict the nearest-neighbour (next-nearest-neighbour) interactions.}
\label{fig1}
\end{figure}  

Among the most famous special cases of the \textit{symmetric (zero-field) eight-vertex model} exhibiting  
the non-universal critical behaviour belongs the spin-1/2 Ising model on two square lattices coupled 
together by the four-spin interaction \cite{Wu71}. As a matter of fact, this system shows the continuously 
varying critical exponents depending merely on the strength of the four-spin interaction \cite{Wan96}. 
Another interesting system, in which the non-universal critical behaviour has precisely been confirmed, 
is being the mixed spin-(1/2, $S$) Ising model on the union jack (centered square) lattice 
\cite{Lip95, Str05}. This mixed-spin system can be examined following the approach originally suggested 
by Lipowski and Horiguchi \cite{Lip95}, who solved one particular case of the mixed spin-(1/2, 1) 
system within the transformation to the corresponding eight-vertex model. Notice that the aforementioned procedure has recently been adapted by one of the present authors to explore another particular case 
of the mixed spin-(1/2, 3/2) union jack lattice \cite{Str05}. Despite the similarity of phase diagrams 
of both these systems, it is somewhat surprising that the mixed spin-(1/2, 1) model exhibits quite different variations of the critical exponents in comparison with its analogous mixed spin-(1/2, 3/2) version. With 
the aim of cumulating a set of exact results that could unambiguously clarify how the critical exponents may depend on the quantum spin number, the main purpose of present work is to provide further extension of the approach used previously in order to bring a deeper insight into the critical behaviour of a general mixed 
spin-(1/2, $S$) Ising model on the union jack lattice with arbitrary quantum spin number $S$.    

This paper is organized as follows. In Section \ref{sec:model}, we provide a detailed formulation of 
the model system and subsequently, the main points of transformation method that ensures an equivalence 
with the eight-vertex model will be briefly mentioned. The most interesting numerical results for a critical behaviour will be presented and particularly discussed in Section \ref{sec:results} for two particular cases of the mixed spin-(1/2, 2) and spin-(1/2, 5/2) systems, respectively. Finally, some concluding remarks are drawn in Section \ref{sec:conclusion}.

\section{Model system and its solution}
\label{sec:model}

Consider the mixed spin-(1/2, $S$) Ising model on the union jack (centered square) 
lattice ${\cal L}$ schematically illustrated in Fig. \ref{fig1}. The mixed-spin union jack lattice 
consists of two interpenetrating sub-lattices ${\cal A}$ and ${\cal B}$, which are formed by the 
spin-1/2 and spin-$S$ sites denoted by empty and filled circles, respectively. The total Hamiltonian 
defined upon the aforedescribed lattice ${\cal L}$ reads:
\begin{eqnarray}
{\mathcal H}_{mix} = - J  \sum_{(m,i) \subset \mathcal J}^{4N} S_{m} \sigma_{i}
     - J' \sum_{(k,l) \subset \mathcal K}^{2N} \sigma_{k} \sigma_{l}
     - D \sum_{m=1}^{N} S_{m}^2,     
\label{HD}
\end{eqnarray}
where $\sigma_i = \pm 1/2$ and $S_m = -S, -S+1, ..., +S$ are Ising spin variables placed on 
the eight- and four-coordinated sites, $J$ denotes the exchange interaction between nearest-neighbouring 
${\cal A}-{\cal B}$ spin pairs and $J'$ labels the interaction between the ${\cal A}-{\cal A}$ spin 
pairs that are next-nearest-neighbours on the union jack lattice ${\cal L}$. Finally, the parameter 
$D$ measures a strength of the uniaxial single-ion anisotropy acting on the spin-$S$ sites and 
$N$ denotes the total number of the spin-1/2 sites. 

\begin{figure}[htb]
\vspace{-0.7cm}
\begin{center}
\includegraphics[width=\textwidth]{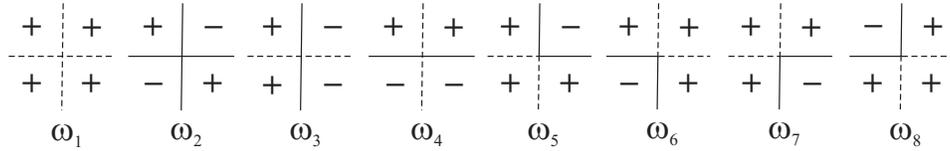}
\end{center}
\vspace{-1.5cm}
\caption{Eight possible spin configurations around each central spin-$S$ site 
and their corresponding line coverings at the vertices of dual square lattice.}
\label{fig2}
\end{figure}

In order to proceed further with calculation, the central spin-$S$ atoms should be firstly decimated from the faces of sub-lattice ${\cal A}$. After the decimation, i.e. after performing a summation over spin states of all spin-$S$ sites, the partition function of the mixed-spin union jack lattice ${\cal L}$ 
can be rewritten as:  
\begin{eqnarray}
{\mathcal Z}_{mix} = \sum_{\{\sigma \}} \prod_{i ,j, k, l} \omega (\sigma_i, \sigma_j, \sigma_k, \sigma_l),    \label{ZD}
\end{eqnarray}
where the summation is performed over all possible spin configurations available on the sub-lattice ${\cal A}$ and the product runs over all $N$ faces (square plaquettes) each composed of one central spin-$S$ site surrounded by four spin-1/2 variables $\sigma_i$, $\sigma_j$, $\sigma_k$, $\sigma_l$ as arranged in Fig.~\ref{fig1}. The Boltzmann factor $\omega (a, b, c, d)$ assigned 
to those faces can be defined as:
\begin{eqnarray}
\omega (a, b, c, d) = \exp[\beta J'(ab + bc + cd + da)/2]  
\sum_{n=-S}^{+S} \exp(\beta D n^2) \cosh[\beta J n (a + b + c + d)],
\label{BF} 
\end{eqnarray}
where $\beta = 1/(k_{\mathrm B} T)$, $k_{\mathrm B}$ is Boltzmann's constant and $T$ stands for 
the absolute temperature. At this stage, the model under investigation can be rather straightforwardly 
mapped onto the eight-vertex model on a dual square lattice ${\cal L_D}$, since the Boltzmann 
factor $\omega (a, b, c, d)$ is being invariant under the reversal of all four spin variables. 
Actually, there are maximally eight different spin arrangements having different energies 
(Boltzmann weights) and these can readily be related to the Boltzmann weights of the eight-vertex 
model on the dual square lattice. If, and only if, the adjacent spins are aligned opposite to each 
other, then solid lines are drawn on the edges of the dual lattice ${\cal L_D}$, otherwise they are 
drawn as broken lines. Diagrammatic representation of eight possible spin arrangements and their 
corresponding line coverings is shown in Fig.~\ref{fig2}.  
It can easily be understood that each of the eight possible line coverings around each vertex of 
the dual lattice corresponds to two spin configurations, one is being obtained from the other by 
reversing all side spins. Since there is even number of solid (broken) lines incident to each vertex of 
the dual lattice ${\cal L_D}$, the model under consideration becomes equivalent with the eight-vertex model. 
With regard to this equivalence, the partition function of the mixed-spin Ising model on the union jack lattice can be expressed in terms of the partition function of the eight-vertex model on the square lattice:
\begin{eqnarray}
{\mathcal Z}_{mix} (T, J, J', D) = 2 {\mathcal Z}_{8-v} (\omega_1, \omega_2, ..., \omega_8).
\label{PF}
\end{eqnarray}   
The factor 2 in this equation comes from the two-to-one mapping between spin and 
vertex configurations (two spin configurations always correspond to one vertex configuration). 

The Boltzmann weights, which correspond to eight possible line coverings of the eight-vertex model 
(Fig. \ref{fig2}), can directly be calculated from equation (\ref{BF}): 
\begin{eqnarray}
\omega_1 \! \! \! &=& \! \! \! 
\exp(\beta J'/2) \sum_{n=-S}^{+S} \exp(\beta D n^2) \cosh(2 \beta J n), \nonumber \\
\omega_2 \! \! \! &=& \! \! \! \exp(-\beta J'/2) \sum_{n=-S}^{+S} \exp(\beta D n^2), \nonumber \\
\omega_3 \! \! \! &=& \! \! \! \omega_4 = \sum_{n=-S}^{+S} \exp(\beta D n^2), \nonumber \\
\omega_5 \! \! \! &=& \! \! \! \omega_6 = \omega_7 = \omega_8 = 
\sum_{n=-S}^{+S} \exp(\beta D n^2) \cosh(\beta J n).  \label{BW}
\end{eqnarray}   
Unfortunately, there does not exist general exact solution for the eight-vertex model with arbitrary 
Boltzmann weights. However, if the weights (\ref{BW}) satisfy the so-called \textit{free-fermion condition}:
\begin{eqnarray}
\omega_1 \omega_2 + \omega_3 \omega_4 = \omega_5 \omega_6 + \omega_7 \omega_8,
\label{FFC}
\end{eqnarray}  
the eight-vertex model then becomes exactly soluble as the \textit{free-fermion model} treated 
several years ago by Fan and Wu \cite{Fan70}. It can readily be proved that the free-fermion 
condition (\ref{FFC}) holds in our case just as $D \to \pm \infty$, or $T \to \infty$. According to this, 
the infinitely strong single-ion anisotropy leads to the familiar phase transitions from the standard 
Ising universality class, since our model system effectively reduces to the simple spin-1/2 Ising model 
on the union jack \cite{Vak66} or the square lattice \cite{Ons44} solved many years ago. Within the 
manifold given by the constraint (\ref{FFC}), the free-fermion model becomes critical as long as:
\begin{eqnarray}
\omega_1 + \omega_2 + \omega_3  + \omega_4 = 2 \mbox{max} \{ \omega_1, \omega_2, \omega_3, \omega_4 \}.
\label{TCFFC}
\end{eqnarray}
It is noteworthy, however, that the critical condition (\ref{TCFFC}) yields rather reliable 
estimate of the criticality within the \textit{free-fermion approximation} \cite{Fan70} 
even if a non-validity of the free-fermion condition (\ref{FFC}) is simply ignored.  

The second branch of exact solution occurs just as the Boltzmann weights (\ref{BW}) satisfy the condition 
of the so-called symmetric (zero-field) eight-vertex model \cite{Bax82}: 
\begin{eqnarray}
\omega_1 = \omega_2, \quad \omega_3 = \omega_4, \quad \omega_5 = \omega_6, \quad \omega_7 = \omega_8.
\label{S8V}
\end{eqnarray} 
As we already have $\omega_3 = \omega_4$, $\omega_5 = \omega_6$, and $\omega_7 = \omega_8$, hence,
the symmetric case is obtained by imposing the condition $\omega_1 = \omega_2$ only, or equivalently:
\begin{eqnarray}
\exp(- \beta J') = \frac{\sum_{n=-S}^{+S} \exp(\beta D n^2) \cosh(2 \beta J n)}
                         {\sum_{n=-S}^{+S} \exp(\beta D n^2)}.
\label{S8V1}
\end{eqnarray}
According to Baxter's exact solution \cite{Bax82} the symmetric model becomes critical on the manifold (\ref{S8V}) if:
\begin{eqnarray}
\omega_1 + \omega_3 + \omega_5  + \omega_7 = 2 \mbox{max} \{ \omega_1, \omega_3, \omega_5, \omega_7 \}.
\label{TCS8V}
\end{eqnarray} 
It is easy to check that $\omega_1$ always represents in our case the greatest Boltzmann weight, thus, 
the condition determining the criticality can also be written in this equivalent form:
\begin{eqnarray}
\Bigl[\sum_{n=-S}^{+S} \exp(\beta_c D n^2) + 2 \sum_{n=-S}^{+S} \! \! \! \! \! && \! \! \! \! \!
\exp(\beta_c D n^2) \cosh(\beta_c J n)\Bigr]^2
\nonumber \\ = \Bigl[ \sum_{n=-S}^{+S} \exp(\beta_c D n^2) \Bigr] \! \! \! \! \! && \! \! \! \! \!
\Bigl[ \sum_{n=-S}^{+S} \exp(\beta_c D n^2) \cosh(2 \beta_c J n) \Bigr],
\label{TCS8V1}
\end{eqnarray}
where $\beta_c = 1/(k_{\mathrm B} T_c)$ and $T_c$ denotes the critical temperature. It should be stressed, nevertheless, that the critical exponents (with exception of $\delta$ and $\eta$) describing a phase transition of the symmetric eight-vertex model depend on the function $\mu = 2 \arctan(\omega_5 \omega_7/ \omega_1 \omega_3)^{1/2}$, in fact:
\begin{eqnarray}
\alpha = \alpha' = 2 - \frac{\pi}{\mu}, \quad \beta = \frac{\pi}{16 \mu}, \quad \nu = \nu' = \frac{\pi}{2 \mu},
\quad \gamma = \frac{7 \pi}{8 \mu}, \quad \delta = 15, \quad \eta = \frac{1}{4}.
\label{CE}
\end{eqnarray} 

\section{Results and discussion}
\label{sec:results}

\begin{figure}[htb]
\vspace{-0.1cm}
\begin{minipage}[t]{0.48\textwidth}
\includegraphics[width=1.05\textwidth]{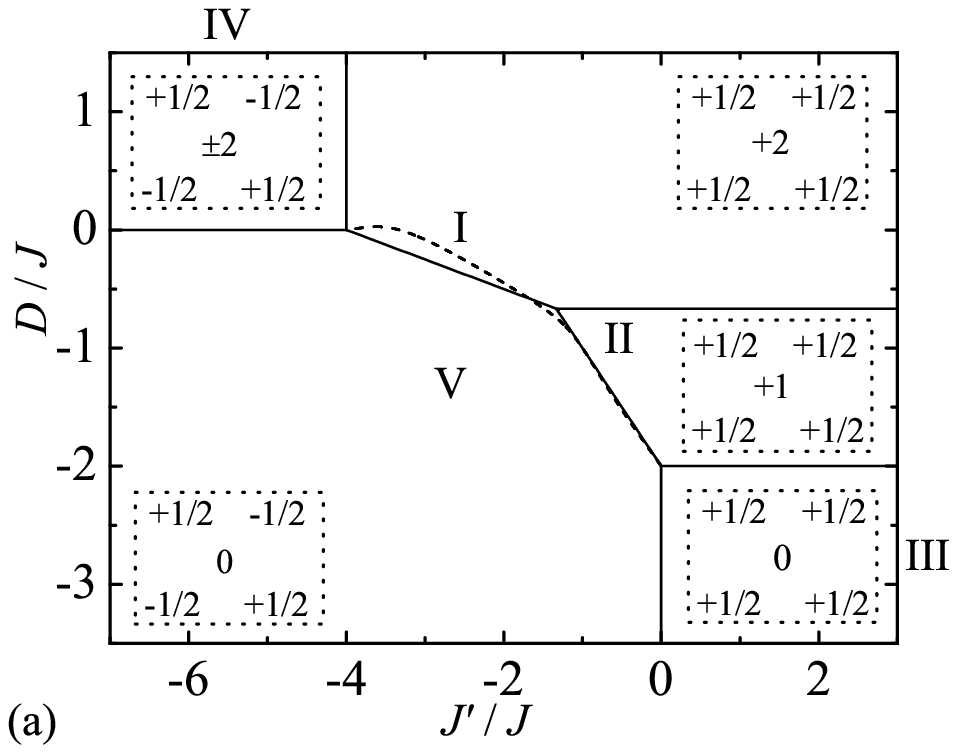}
\end{minipage}
\vspace{-0.1cm}
\begin{minipage}[t]{0.48\textwidth}
\includegraphics[width=1.05\textwidth]{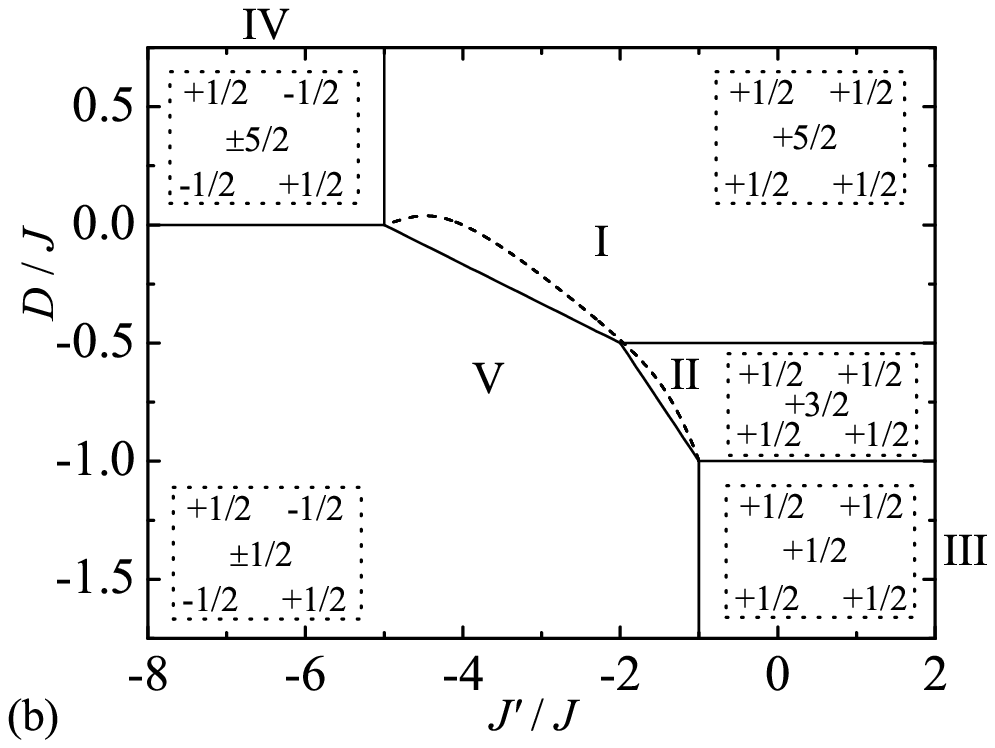}
\end{minipage}
\vspace{-0.1cm}
\caption{Ground-state phase diagram of the mixed spin-(1/2, $S$) Ising model on the union jack
lattice for: (a) $S = 2$, (b) $S = 5/2$. Dotted rectangles schematically illustrate typical spin configurations within the basic unit cell of each phase, broken line connecting triple points 
shows a projection of the critical line (\ref{TCS8V1}) into $J'-D$ plane.}
\label{fig3}
\end{figure}

Before proceeding to a discussion of the most interesting results obtained for critical properties  
of two particular mixed spin-(1/2, 2) and spin-(1/2, 5/2) systems on the union jack lattice, 
let us briefly comment on their ground-state phase diagrams. Solid lines displayed in Fig. \ref{fig3} 
represent ground-state phase boundaries separating five different long-range ordered phases, which 
emerge in the ground state when $J > 0$. Spin order drawn in dotted rectangles shows a typical 
spin configuration within the basic unit cell of each phase. It is quite obvious from Fig. \ref{fig3}
that both investigated spin systems have rather similar ground-state phase diagrams. Actually,
when moving along the vertical axis in the up-down direction, then the single-ion anisotropy parameter
$D/J$ forces the central spins to lower their spin state, while when moving along the horizontal axis
in the right-left direction, then the sufficiently strong antiferromagnetic next-nearest-neighbour interaction 
$J'$ alters the structure of the ground state due to a competition with the nearest-neighbour interaction $J$. It is noticeable that the central spin-$S$ atoms are free to flip within the phases IV and V, 
where the interaction energy between spin-$S$ and spin-1/2 sites effectively cancels out as a result 
of the antiferromagnetic alignment between the latter spins. Accordingly, the phases IV and V 
exhibit a remarkable combination of the spin order on the sub-lattice ${\cal A}$ (spin-1/2 sites) 
and the disorder that appears on the sub-lattice ${\cal B}$ (spin-$S$ sites). There is nevertheless 
one principal difference between the two investigated spin systems: the central spins are non-degenerate 
within the phase V whenever the mixed spin-(1/2, $S$) system consists of the integer-value spin-$S$ atoms 
as they reside their lowest 'non-magnetic' spin state $S = 0$. Furthermore, it is also quite illuminating 
to see from Fig. \ref{fig3} that there exist direct phase transitions between the phases I-V and II-V 
in the range of moderate values of $D/J$ and $J'/J$, which are accompanied by a change 
of the spin order at the sub-lattice ${\cal A}$ as well as ${\cal B}$.

Next, our particular interest will be devoted to a broken line that depicts in Fig. \ref{fig3} a projection 
of the exact critical line (\ref{TCS8V1}) into the $J'-D$ plane. As it can be clearly seen, this projection crosses $T=0$ plane along the first-order transition lines between the phases I-V and II-V, respectively. Referring to the studies reported on previously, this remarkable line of critical points can be identified as a line of bicritical points that have the non-universal continuously varying critical exponents 
\cite{Lip95, Str05}. It will be shown later on that there indeed meet two second-order phase transition lines with one first-order transition line at each bicritical point and altogether, the line of bicritical points bounds a coexistence surface between different ordered phases. In this respect, we may conclude that the special critical line consisting of bicritical points determines a location of phase transitions between the phases I-V and II-V, respectively. 

\begin{figure}[htb]
\vspace{-0.1cm}
\begin{minipage}[t]{0.48\textwidth}
\includegraphics[width=1.05\textwidth]{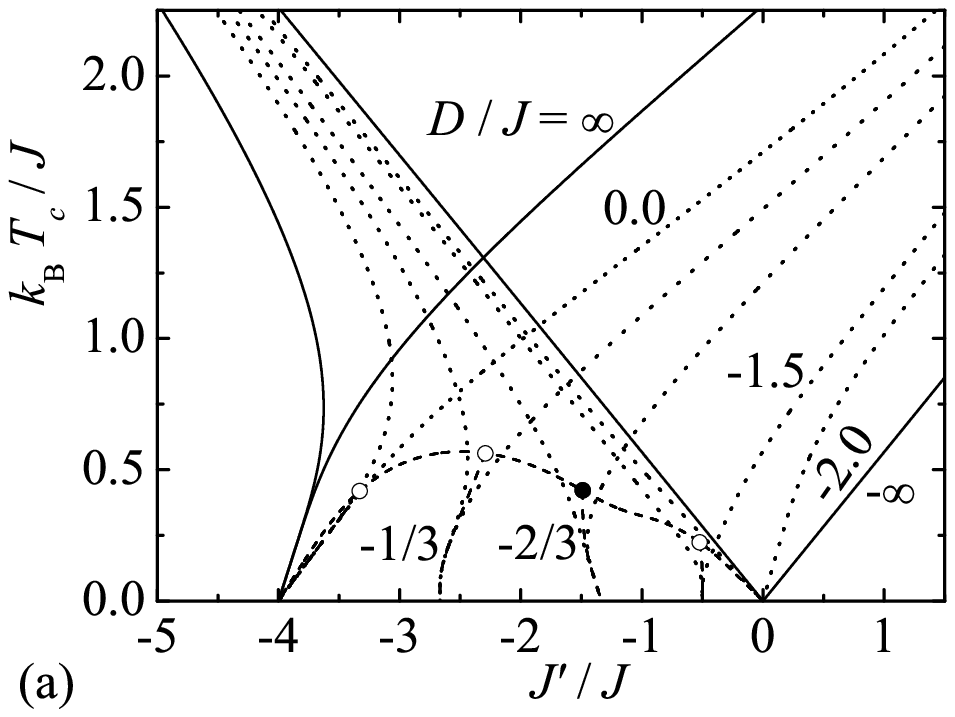}
\end{minipage}
\vspace{-0.1cm}
\begin{minipage}[t]{0.48\textwidth}
\includegraphics[width=1.05\textwidth]{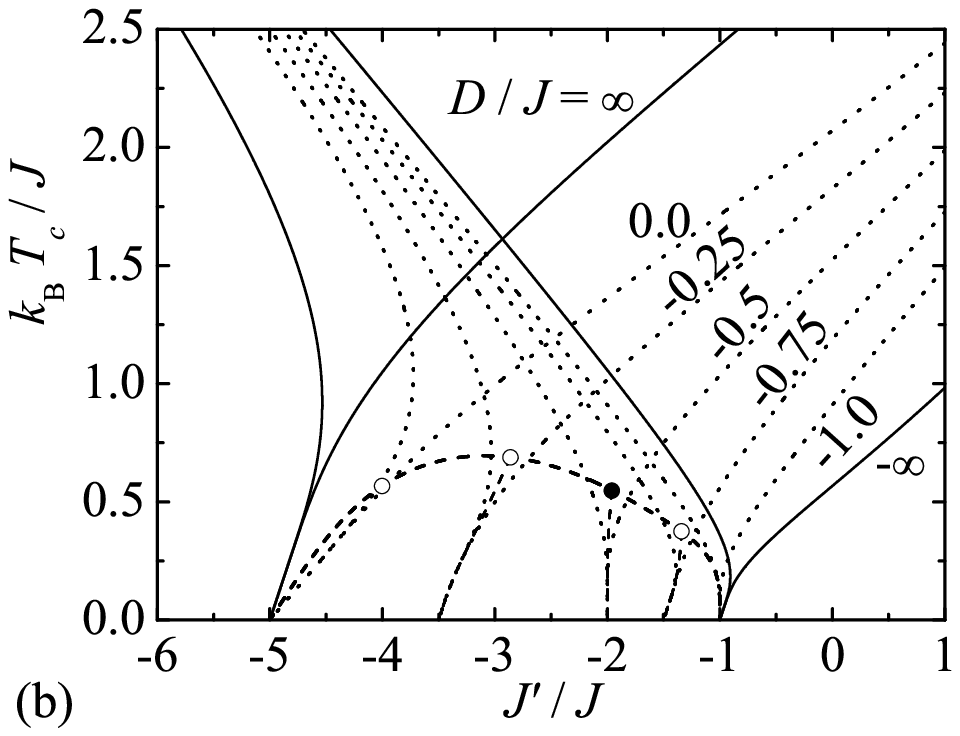}
\end{minipage}
\vspace{-0.1cm}
\caption{Phase boundaries of the mixed spin-(1/2, $S$) Ising model 
on the union jack lattice for: (a) $S = 2$, (b) $S = 5/2$. For details see the text.}
\label{fig4}
\end{figure}

Now, let us construct a global finite-temperature phase diagram. For this purpose, Fig.~\ref{fig4}
displays the critical temperature as a function of the ratio $J'/J$ for several values of the single-ion anisotropy strengths $D/J$. Critical boundaries depicted as solid lines represent exact critical lines obtained from the free-fermion solution (\ref{TCFFC}), which can be applied on the variety (\ref{FFC}) 
fulfilled in two limiting cases $D/J \to \pm \infty$. On the other hand, dotted critical lines show estimated critical temperatures calculated from the free-fermion approximation simply ignoring a non-validity of the 
free-fermion condition (\ref{FFC}) for any finite value of $D/J$. Second branch of exact solution, 
which is related to the aforementioned critical line of the symmetric (zero-field) eight-vertex model (\ref{TCS8V1}) on the variety (\ref{S8V1}), is displayed as a round broken line. Broken lines terminating 
at circled points show for completeness the zero-field condition (\ref{S8V1}) calculated for several 
values $D/J$, since the zero-field condition should be identical to the first-order transitions 
terminating at bicritical points. As a matter of fact, there are strong indications supporting this 
concept: the zero-field condition (\ref{S8V1}) should always show a coexistence of different phases 
and also in our case it always starts from and ends up on points giving the coexistence 
of the phases I-V and II-V, respectively. 

By combining the results illustrated in Figs. \ref{fig3}(a) and \ref{fig4}(a) we may come to 
the following conclusions: if the central spins are occupied by spin-2 atoms, then a right 
(left) wing of the displayed critical boundaries corresponds to the phase I (IV) if $D/J > 0$, 
while it corresponds to the phase III (V) if $D/J < -2$. Within the range $-2 < D/J < 0$ the 
right wing always corresponds to the phase V, but the left wing corresponds either to the phase 
I or II depending on whether $D/J$ is greater or smaller than the boundary value $D_{\rm b}/J = -2/3$. 
The situation exactly at this boundary value is also depicted in Fig. \ref{fig4}(a), 
three critical lines should terminate at the special bicritical point (filled circle) in which 
three different phases I, II and V coexist together. The area 
bounded by the line of bicritical points and the zero-field condition for $D_{\rm b}/J$ 
consequently separates two different regions of coexistence: the area on the lhs (rhs) from the 
zero-field condition determines a coexistence of the phases I (II) and V, respectively. 
The first-order transition line between the phases I (II) and V is shown for one particular case
$D/J = -1/3$ ($D/J = -3/2$). Another interesting finding follows from a detailed comparison of 
Figs. \ref{fig3}(a) and \ref{fig4}(a): in addition to the finite-temperature bicritical point 
(shown as empty circle) there appears another bicritical point which is located precisely at zero 
temperature when $D/J = 0$ is selected. Notice that two finite-temperature bicritical points may even 
occur when $D/J$ is slightly above zero, however, upon strengthening of $D/J$ they move, along the line 
of bicritical points, closer to each other until they coalesce at $D_{\rm max}/J = 0.0267$.

Similar finite-temperature phase diagram as discussed formerly for the mixed spin-(1/2, 2) Ising model 
on the union jack lattice exhibits also its analogous mixed spin-(1/2, 5/2) version (compare Figs. \ref{fig4}(a) and (b)). For the sake of brevity, we shall therefore not repeat its adequate description 
here and we just quote the most essential differences. Of course, the most interesting region of criticality is now restricted to another range of single-ion anisotropy where the phase coexistence emerges 
($-1 < D/J < 0$). It is noteworthy that the midpoint of this interval ($D_{\rm b}/J = -0.5$) now determines 
a special point in which the coexistence of three different phases (I, II and V) occurs. Further, the 
doubling of bicritical points at fixed value of $D/J$ is even more pronounced for the mixed spin-(1/2, 5/2) system, actually, the coalescence of bicritical points occurs just as $D_{\rm max}/J = 0.0385$ is approached.
Apart from these trivial findings, the most significant difference can be observed in the exact 
critical line calculated in the limit $D/J \to - \infty$. On the one hand, the reduced critical temperature 
($k_{\rm B} T_c / J$) of the mixed spin-(1/2, 2) system increases linearly with $J'/J$, what means, 
that the critical temperature is independent of $J$ and depends merely on $J'$. 
It should be realized that both phases III and V stable in the limit $D/J \to - \infty$
possess the central spin-2 atoms in their 'non-magnetic' 
spin state $S=0$ and owing to this fact, the model system with integer spin-$S$ atoms effectively 
behaves as a simple spin-1/2 square lattice. A slope of the $T_c = f(J')$ dependence indeed gives constant 
values $k_{\rm B} T_c / J' = \pm [2 \ln (1 + \sqrt{2})]^{-1}$ for the phase III (upper sign) 
and V (lower sign) in accord with an exact critical point of the spin-1/2 ferromagnetic 
(antiferromagnetic) Ising square lattice \cite{Ons44}. On the other hand, 
the critical temperature of the mixed spin-(1/2, 5/2) system taken in the limit $D/J \to - \infty$
is strongly reminiscent to the one at $D/J \to \infty$. Namely, both the dependences 
resemble the same critical line of the spin-1/2 Ising model on the union jack lattice \cite{Vak66}; 
the dependence calculated for $D/J \to \infty$ is just rescaled and shifted by a factor of five 
due to five-times higher spin value of the central spins. 
Finally, we should remark a feasible appearance of the reentrant phase transitions 
observable in both the investigated spin systems. It is quite apparent that the reentrance appears 
on account of the competition between the nearest- and next-nearest-neighbour interactions, in fact, 
both the higher-temperature ordered phases IV and V exhibit the coexistence of a partial order 
and disorder, which has been conjectured as a necessary condition for appearance of the 
reentrant transitions \cite{Die04}. 

\begin{figure}[htb]
\begin{minipage}[t]{0.48\textwidth}
\includegraphics[width=1.05\textwidth]{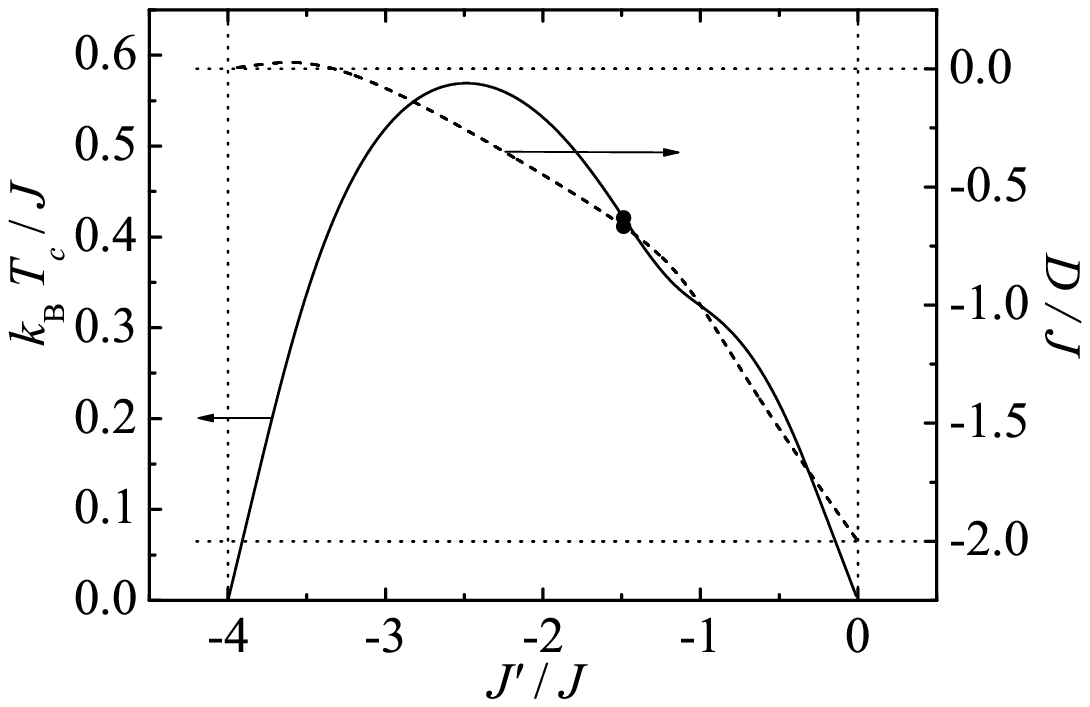}
\vspace{-0.9cm}
\caption{Dependence scaled to a left axis shows how the critical temperature changes with 
the ratio $J'/J$, the curve scaled with respect to a right axis depicts variation of $D/J$ along this line.
Dotted lines are guides for eyes.}
\label{fig5}
\end{minipage}
\hfil
\begin{minipage}[t]{0.48\textwidth}
\includegraphics[width=1.05\textwidth]{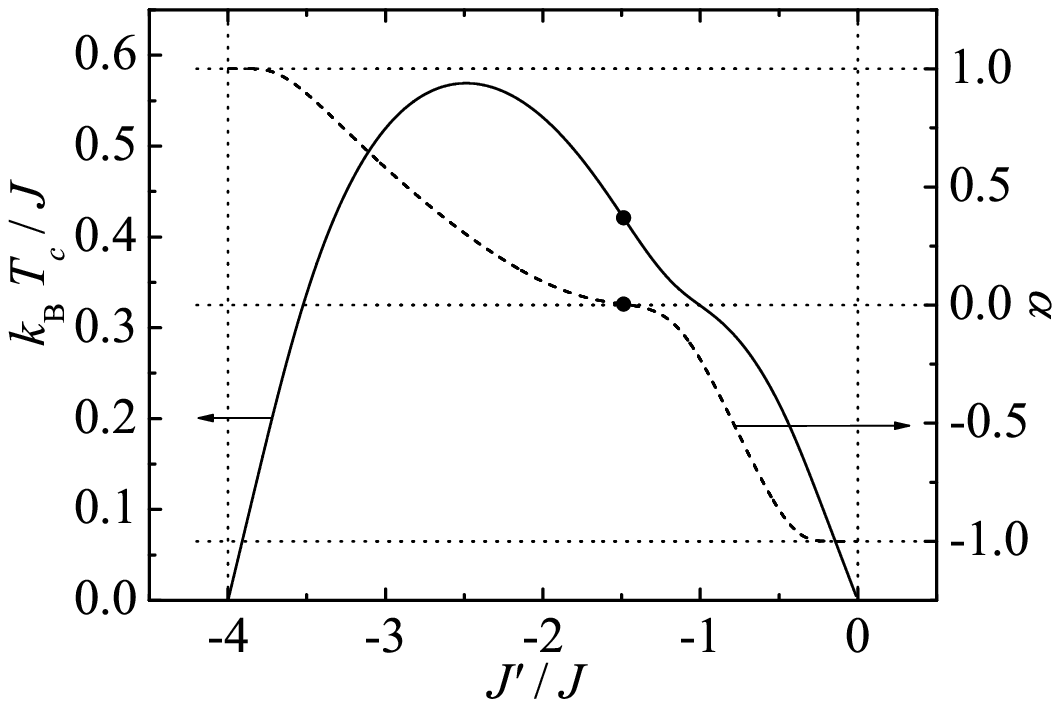}
\vspace{-0.9cm}
\caption{Dependence scaled to a left axis shows how the critical temperature changes with the ratio $J'/J$, the curve scaled with respect to a right axis depicts variations of the critical exponent $\alpha$ along this line.}
\label{fig6}
\end{minipage}
\end{figure}

Last, let us take a closer look at the variations of critical exponents along the line 
of bicritical points. At first, we shall investigate the line of bicritical points for the 
mixed spin-(1/2, 2) system. Fig. \ref{fig5} displays a projection of this critical line 
\vspace{-0.75cm}
\begin{figure}[htb]
\begin{minipage}[t]{0.48\textwidth}
\includegraphics[width=1.05\textwidth]{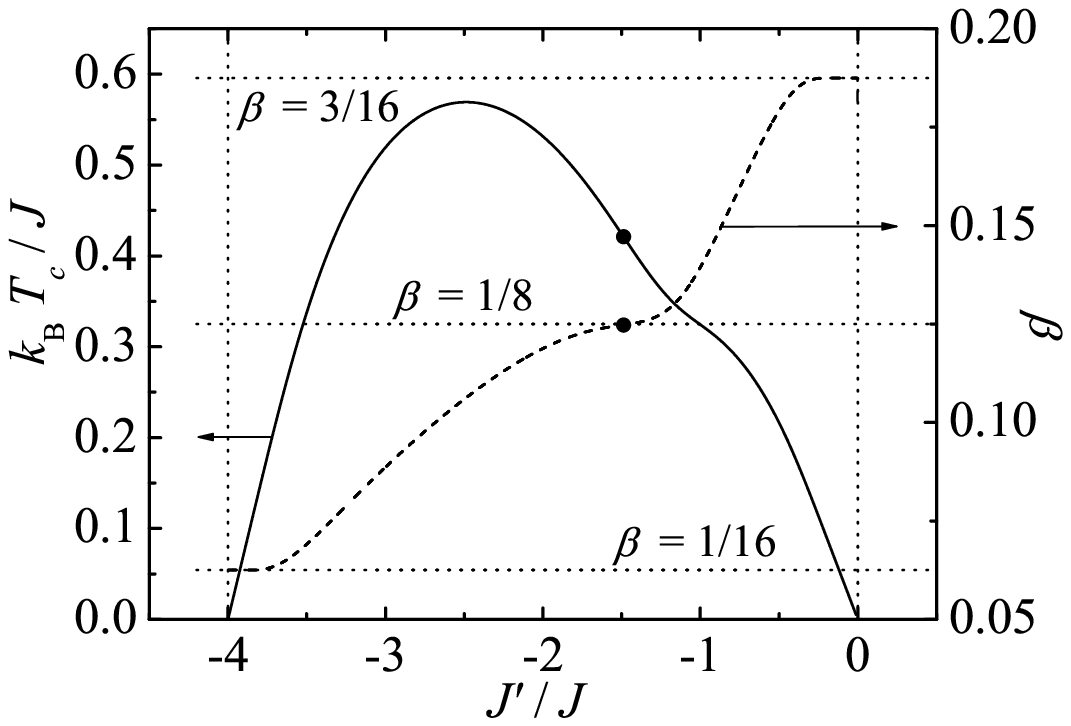}
\vspace{-0.9cm}
\caption{The same as in Fig. \ref{fig6}, but the critical exponent $\beta$ is now scaled 
with respect to the right axis.}
\label{fig7}
\end{minipage}
\hfil
\begin{minipage}[t]{0.48\textwidth}
\includegraphics[width=1.05\textwidth]{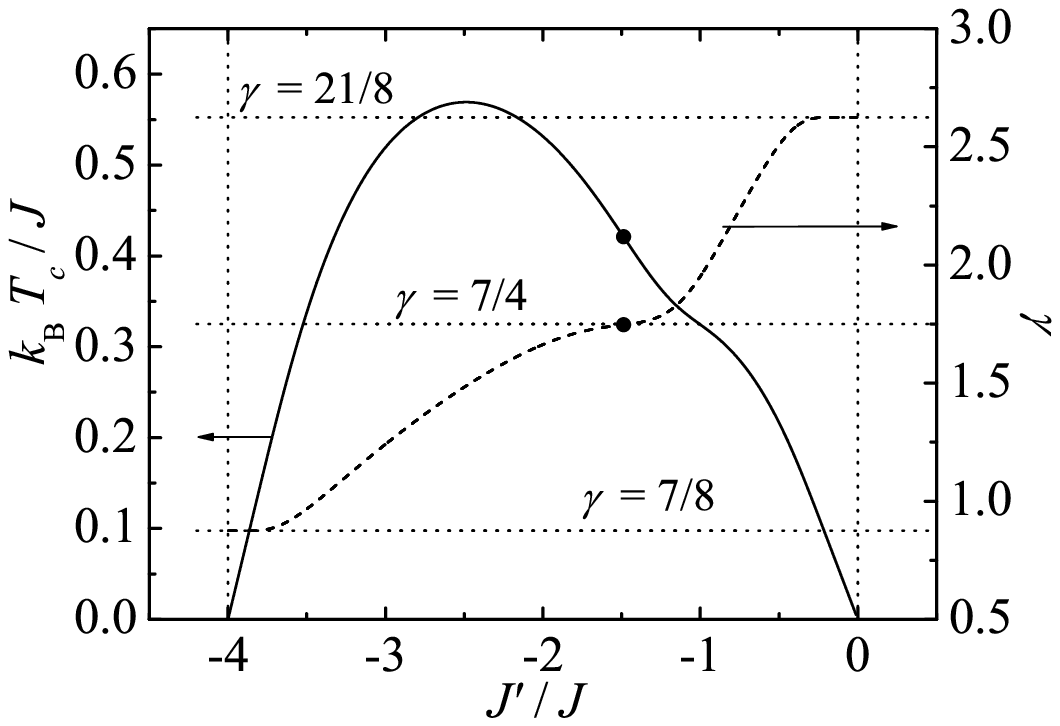}
\vspace{-0.9cm}
\caption{The same as in Fig. \ref{fig6}, but the critical exponent $\gamma$ is now scaled 
with respect to the right axis.}
\label{fig8}
\end{minipage}
\end{figure}
into the $J'-T_c$ plane (the dependence scaled to the left axis) and simultaneously, 
its projection into the $J'-D$ plane which is scaled with respect to the right axis. 
Next, Figs. \ref{fig6}, \ref{fig7} and \ref{fig8} show how the critical indices $\alpha$, 
$\beta$, and $\gamma$, respectively, change along this critical line. It is quite evident
from these figures that the exponents $\beta$ and $\gamma$ monotonically increase upon 
strengthening of $J'/J$, while the critical exponent $\alpha$ shows a monotonous decrease 
as $J'/J$ increases. Another interesting fact to observe here is that the critical exponents 
of special bicritical point (filled circle), at which the coexistence of three phases 
I, II and V occurs, are close (but not exactly equal) to their universal values.
Finally, it is also worthwhile to mention that the bicritical points of the mixed spin-(1/2, 2) 
system exhibit a rather large variations of critical exponents ranging: 
$\alpha \in \langle -1, 1 \rangle$, $\beta \in \langle 1/16, 3/16 \rangle$ and 
$\gamma \in \langle 7/8, 21/8 \rangle$.   

\begin{figure}[htb]
\begin{minipage}[t]{0.48\textwidth}
\includegraphics[width=1.05\textwidth]{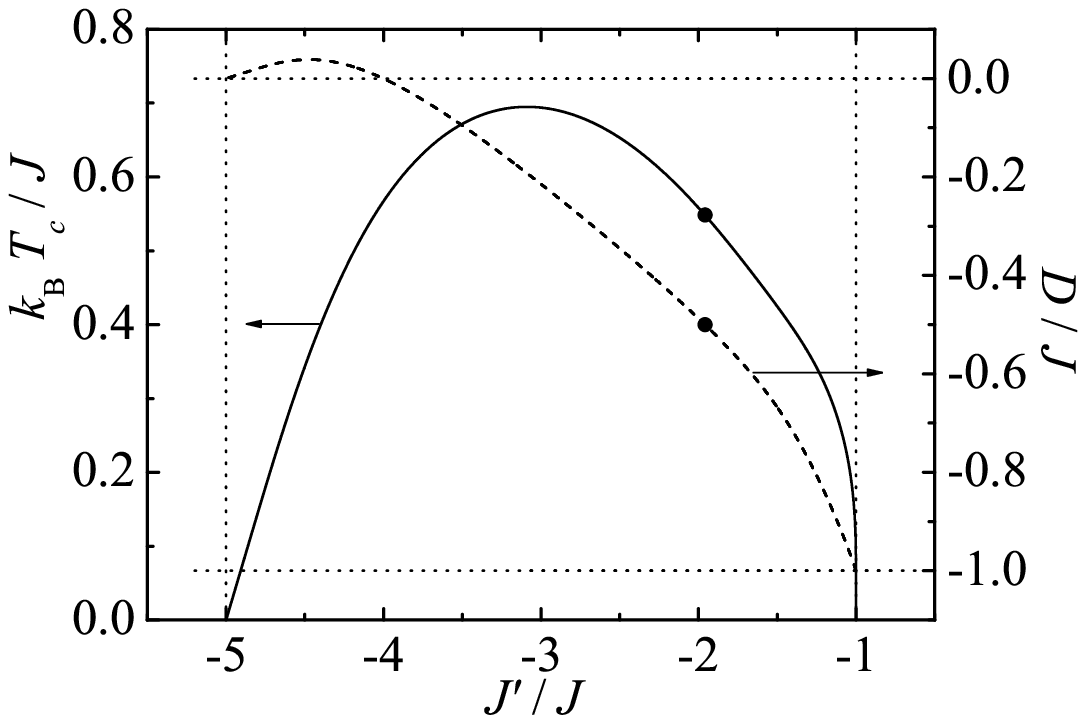}
\vspace{-0.9cm}
\caption{Dependence scaled to a left axis shows how the critical temperature changes with 
the ratio $J'/J$, the curve scaled with respect to a right axis depicts variation of $D/J$ along this line.
Dotted lines are guides for eyes.}
\label{fig9}
\end{minipage}
\hfil
\begin{minipage}[t]{0.48\textwidth}
\includegraphics[width=1.05\textwidth]{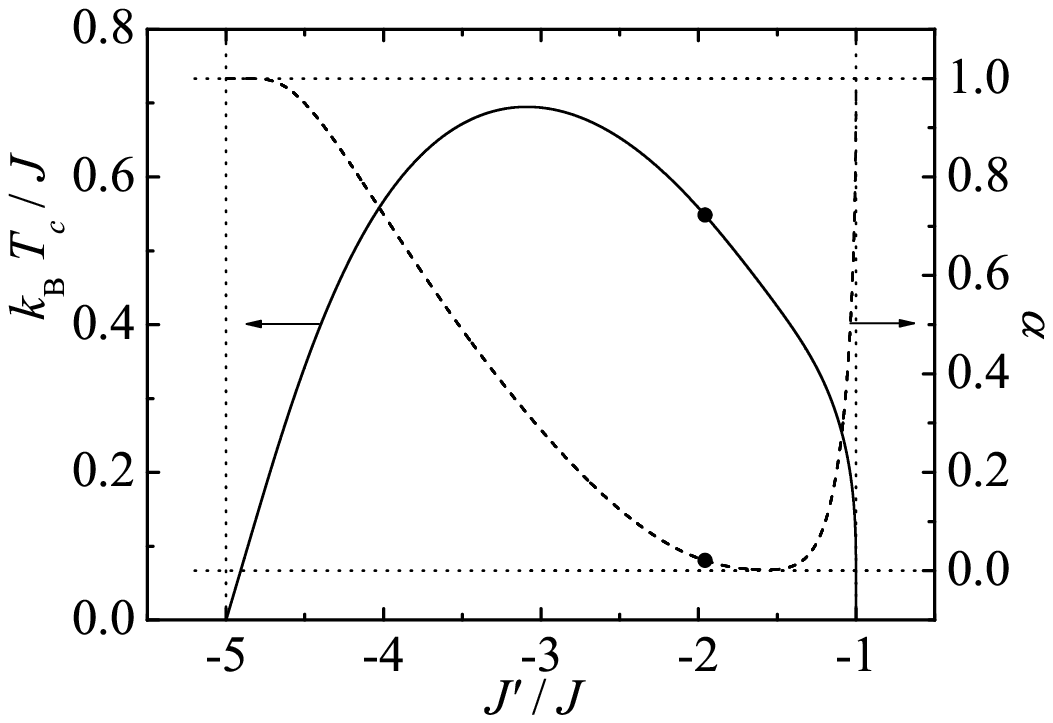}
\vspace{-0.9cm}
\caption{Dependence scaled to a left axis shows how the critical temperature changes with the ratio $J'/J$, the curve scaled with respect to a right axis depicts variations of the critical exponent $\alpha$ along this line.}
\label{fig10}
\end{minipage}
\end{figure}

\begin{figure}[htb]
\begin{minipage}[t]{0.48\textwidth}
\includegraphics[width=1.05\textwidth]{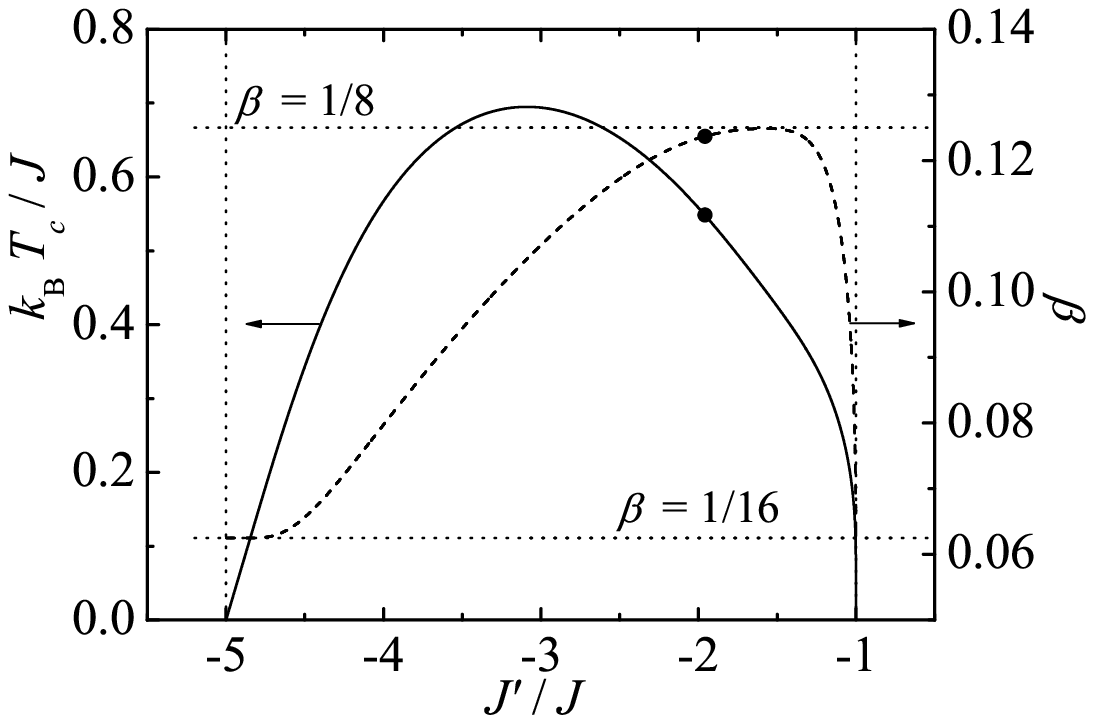}
\vspace{-0.9cm}
\caption{The same as in Fig. \ref{fig10}, but the critical exponent $\beta$ is now scaled 
with respect to the right axis.}
\label{fig11}
\end{minipage}
\hfil
\begin{minipage}[t]{0.48\textwidth}
\includegraphics[width=1.05\textwidth]{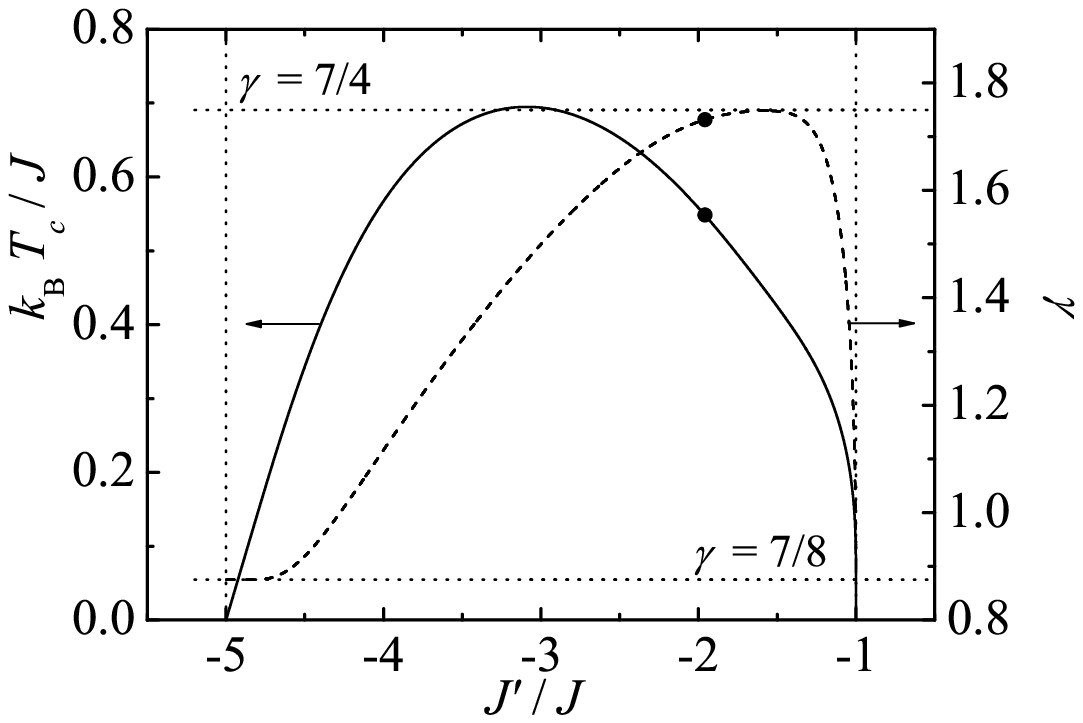}
\vspace{-0.9cm}
\caption{The same as in Fig. \ref{fig10}, but the critical exponent $\beta$ is now scaled 
with respect to the right axis.}
\label{fig12}
\end{minipage}
\end{figure}

For comparison, Figs. \ref{fig9}-\ref{fig12} illustrate relevant dependences of the critical 
temperature and critical exponents for the mixed spin-(1/2, 5/2) system. As it can be clearly 
seen, the critical exponents $\beta$ and $\gamma$ approach their smallest possible 
values $1/16$ and $7/8$ by reaching both triple points with zero critical temperature, while the 
critical index $\alpha$ approaches there its greatest possible value $1$. With high certainty, 
the same critical exponents of both triple points can be related to a similar spin alignment of 
three coexisting phases I-II-V and III-IV-V. As a matter of fact, this conjecture is in a good accord 
with what is observed in the mixed spin-(1/2, $S$) system with integer spin-$S$ atoms, where the 
critical indices at both triple points differ significantly due to a presence of 'non-magnetic' 
state $S=0$ energetically favored in the phases III and V. It is worthy to notice, moreover, that 
the greatest values of critical indices $\beta$ and $\gamma$, as well as, the smallest value of critical exponent $\alpha$, are being almost completely equal to the values predicted by the universality 
hypothesis for planar Ising systems. 
\vspace{-0.75cm}
With all this in mind, we may conclude that the mixed spin-(1/2, $S$) systems with the half-odd-integer 
spin-$S$ atoms exhibit less variance in the changes of critical exponents as they are restricted 
to smaller ranges: 
$\alpha \in \langle 0, 1 \rangle$, $\beta \in \langle 1/16, 1/8 \rangle$ and 
$\gamma \in \langle 7/8, 7/4 \rangle$.

\section{Concluding Remarks}
\label{sec:conclusion}

The present article deals with a critical behaviour of the mixed spin-(1/2, $S$) Ising model on the 
union jack lattice that represents one of few exactly soluble planar Ising systems accounting also 
for the interaction beyond nearest neighbours. It is worthwhile to remark that the relatively precise 
and concise information on criticality of the considered model system has been obtained through 
an establishment of the mapping relationship with its corresponding eight-vertex model, which has 
exactly been solved in some particular cases when its Boltzmann weights satisfy either the free-fermion 
condition (\ref{FFC}), or the zero-field condition (\ref{S8V}). In agreement with our expectations, an interplay between the nearest-neighbour interaction, the competing next-nearest-neighbour interaction and the single-ion anisotropy gives rise to a rather complex critical behaviour displayed in the reentrant phase transitions, the weak universal critical behaviour, a presence of first- as well as second-order transitions 
and so on. 

The most challenging question concerning with the critical behaviour of the considered model system should 
be connected with the remarkable line of bicritical points, which bounds a coexistence surface between 
phases with different spin alignment. It is noteworthy that the results presented in the pioneering 
work of Lipowski and Horiguchi \cite{Lip95} imply two times greater variations of critical exponents
along the line of bicritical points for the mixed spin-(1/2, 1) union jack lattice when comparing them 
with the results recently reported for its analogous mixed spin-(1/2, 3/2) version \cite{Str05}. 
Owing to this fact, the main aim of the present work was to clarify how the critical exponents may 
depend on the quantum spin number $S$. In view of the exact results presented here it is quite tempting 
to conjecture that the mixed spin-(1/2, $S$) models may exhibit different variations of critical exponents depending on whether spin-$S$ sites are occupied by the half-odd-integer spins, or the integer ones. 
Of these systems, the mixed-spin systems with the integer spin-$S$ atoms have been found to exhibit 
greater changes of the critical exponents. 

\begin{acknowledgement}
The authors acknowledge financial support of this work provided under the scientific grants 
VVGS 12/2006, VEGA 1/2009/05 and APVT 20-005204.
\end{acknowledgement}

\end{document}